\documentclass[aps,preprint,amssymb]{revtex4}
\usepackage{graphicx}
\usepackage{dcolumn}
\usepackage{amsmath}
\usepackage{amssymb}
\usepackage{subfigure,amsmath,verbatim,moreverb,bm}

\def\xv{{\bf x}}

\def\jv{{\bf j}}

\def\Av{{\bf A}}
\def\Bv{{\bf B}}

\def\vv{{\bf v}}

\def\xiv{\bm{\xi}}
\def\Acal{{\mathcal A}}
\def\Vcal{{\mathcal V}}
\def\Acalv{{\bm{\mathcal A}}}

\begin{document}
\title{A unified approach to the density-potential mapping in a family of time-dependent density functional theories}
\author{I. V. Tokatly}
\email{Ilya_Tokatly@ehu.es}
\affiliation{Nano-Bio Spectroscopy group and ETSF Scientific Development Centre, 
  Departamento de F\'isica de Materiales, Universidad del Pa\'is Vasco UPV/EHU, E-20018 San Sebasti\'an, Spain}
\affiliation{IKERBASQUE, Basque Foundation for Science, E-48011 Bilbao, Spain}

\date{\today}
\begin{abstract}
It is shown that the density-potential mapping and the ${\cal V}$-representability problems in the time-dependent current density functional theory (TDCDFT) are reduced to the solution of a certain many-body nonlinear Schr\"odinger equation (NLSE). The derived NLSE for TDCDFT adds a link which bridges the earlier NLSE-based formulations of the time-dependent deformation functional theory (TDDefFT) and the time-dependent density functional theory (TDDFT). We establish close relations between the nonlinear many-body problems which control the existence of TDCDFT, TDDFT, and TDDefFT, and thus develop a unified point of view on the whole family of the TDDFT-like theories.
\end{abstract}
\maketitle

\section{Introduction}
An accurate description of quantum many-electron dynamics is one of the most challenging problems in the modern theoretical physics. Because a direct application of the standard many-body techniques to interacting nonequilibrium systems becomes prohibitively complicated, the time-dependent density functional theory (TDDFT)\cite{TDDFT2006} remains practically the only available tool for modelling dynamics of realistic systems at a reasonable computational cost. By formally avoiding a full solution of the complicated many-body problem, TDDFT gives a direct access to experimentally relevant observables, such as the density of particles or the density of current. Within the most common Kohn-Sham (KS) implementation of TDDFT the exact density is calculated by propagating mean-field-like equations of motion for a fictitious system of noninteracting KS particles, which enormously simplifies the problem. Therefore a growing popularity of TDDFT is absolutely not surprising (see, e.~g., recent special issues of PCCP \cite{TDDFT-PCCP} and TEOCHEM \cite{TDDFT-TEOCHEM}, as well as this volume).

Practical applications of TDDFT to realistic many-electron systems unavoidably employ approximations. However, in almost every paper, where TDDFT is applied, it is introduced as a formally exact theory. Unfortunately, at the present stage, the common statements of exactness express our expectations/believes more than the actual situation in the field of mathematical foundations of TDDFT and related theories. At the formal level TDDFT is based on two main statements: (i) A map from the external potential to the density is unique and invertable; (ii) A given density can be obtained from the Schr\"odinger dynamics driven by a properly chosen external potential (for example, the density coming from the interacting system can be reproduced by adjusting a potential in a fictitious noninteracting system). The first statement is known as the Runge-Gross mapping theorem \cite{RunGro1984}. The statement (ii) is commonly referred to as a ${\cal V}$-representability problem, which is a corner stone of the KS formalism. 

A disturbing fact is that the Runge-Gross proof \cite{RunGro1984} of the mapping theorem is only valid for potentials which are analytic functions of time ($t$-analytic) and can be represented around the initial time by a convergent Taylor series. The conditions of the ${\cal V}$-representability theorem, as formulated by van~Leeuwen, \cite{vanLeeuwen1999} are even more restrictive. One also assumes the $t$-analyticity of the density to uniquely recover coefficients in the power series expansion of the potential supporting that density, while a uniform convergence of the recovered series is taken for granted. Therefore our present justification of TDDFT and its generalizations, such as time-dependent current density functional theory (TDCDFT) \cite{GhoDha1988,Vignale2004}, is limited to a very narrow class of external potentials. It should be emphasized that the $t$-analyticity of the potential, and even its infinite smoothness in space, are not sufficient to guarantee the $t$-analyticity of the density \cite{Maitra2010} required for the present formulation of the ${\cal V}$-representability theorem \cite{vanLeeuwen1999}. Hence, strictly speaking, the precise conditions of ${\cal V}$-representability are still unknown. Obviously, a better understanding of the fundamental issues of TDDFT is highly desirable, especially in a view of explosively growing number of practical applications.

A commonly accepted weak point of the existing proofs both in TDDFT \cite{RunGro1984,vanLeeuwen1999} and in TDCDFT \cite{GhoDha1988,Vignale2004} is that the mathematical statement of the problem explicitly relies on the power series representation of the densities and/or potentials. It is absolutely clear that no further progress is possible without finding a new formulation, which does not require, at least in principle, the power series expansion of the basic quantities. In several recent works \cite{TokatlyPRB2007,TokatlyPCCP2009,Maitra2010,Tokatly2010arxiv,RugLee2010arxiv} such a new route to the proof of existence of TDDFT-like theories has been proposed. Namely, the density-potential mapping and the $\Vcal$-representability problems can be restated in the form of the existence and uniqueness of solutions to a certain universal nonlinear Schr\"odinger equation (NLSE). The universal NLSE has been first formulated in the context of the time-dependent deformation functional theory (TDDefFT)\cite{TokatlyPRB2007,TokatlyPCCP2009}, where it appears as a natural step in solving the many-body problem in the comoving reference frame. In Ref.~\cite{Maitra2010} a conceptually similar formulation has been proposed for TDDFT, and more recently the NLSE approach has been adapted to rigorously prove the uniqueness and existence theorems for a lattice version of TDCDFT \cite{Tokatly2010arxiv}.

Despite the NLSE formulations proposed for TDDefFT in \cite{TokatlyPRB2007,TokatlyPCCP2009} and for TDDFT in \cite{Maitra2010} are similar conceptually, their formal realizations look very different and appear to be completely disconnected. One of the aims of the present paper is find a relation between these two theories. I will derive the NLSE formulation of TDCDFT and show that it is precisely the missing link which bridges the two abovementioned proposals. We will see that the concept of a universal many-body NLSE provides us with a unified point of view on the whole family of TDDFT-like formalisms, namely, TDCDFT, TDDefFT, and TDDFT itself. This unification is the main result of the present work.

The structure of the paper is the following. In Sec.~IIA I formulate the NLSE approach to the mapping and the $\Vcal$-representability problems in TDCDFT. An equivalence of the derived NLSE for TDCDFT and the NLSE for TDDefFT \cite{TokatlyPRB2007,TokatlyPCCP2009} is proven in  Sec.~IIB. In Sec.~IIC I establish a connection of the NLSE formalism to the Vignale's approach to the TDCDFT mapping theorem \cite{Vignale2004}. In Sec.~III I derive a new NLSE formulation of TDDFT, and demonstrate its connection, on one hand, to the TDCDFT-NLSE of Sec.~II, and, on the other hand to TDDFT-NLSE proposed in Ref.~\cite{Maitra2010}. This uncovers interrelations between all members of the TDDFT-family and completes the unification.
The main results of the paper are summarized in Conclusion.

\section{Time-dependent current density functional theory}

\subsection{Nonlinear inverse many-body problem in TDCDFT}

Let us consider a system of $N$ identical particles in the presence of time dependent external scalar $U(\xv,t)$ and vector ${\bf A}(\xv,t)$ potentials. The corresponding many-body wave function 
$\Psi(\xv_{1},\dots,\xv_{N},t)$
is a solution to the time-dependent Schr\"odinger equation
\begin{equation}
 i\partial_t\Psi(\xv_{1},\dots,\xv_{N},t) = H\Psi(\xv_{1},\dots,\xv_{N},t)
\label{SE}
\end{equation}
with the following Hamiltonian 
\begin{equation} 
H = \sum_{j = 1}^{N}\left[\frac{(-i\nabla_{j} 
- {\bf A}(\xv_{j},t))^{2}}{2m} + U(\xv_{j},t)\right] +
\frac{1}{2}\sum_{j\ne k} V(|\xv_{j}-\xv_{k}|)
\label{H}
\end{equation}
where $\nabla_{j}=\partial/\partial\xv_j$, and $V(|\xv-\xv'|)$ is the interaction potential. For a given initial condition, 
\begin{equation}
\Psi(\xv_{1},\dots,\xv_{N},0) = \Psi_0(\xv_{1},\dots,\xv_{N}),
 \label{InitialPsi}
\end{equation}
the dynamics of the system is completely specified by Eq.~(\ref{SE}). 

The Schr\"odinger equation (\ref{SE}) is invariant under the following gauge transformation
\begin{eqnarray}
 \label{Psi-gauge}
\Psi(t)&\to& e^{-i\sum_{j}\chi(\xv_{j},t)}\Psi(t),\\
\label{A-gauge}
{\bf A}(\xv,t) &\to& {\bf A}(\xv,t) + \nabla\chi(\xv,t), \\
\label{U-gauge}
U(\xv,t) &\to& U(\xv,t) - \partial_t \chi(\xv,t)
\end{eqnarray}
Because the physical results are independent of the gauge we are free to choose a gauge function $\chi(\xv,t)$ which is most suitable for a particular problem. We will see that TDCDFT is most elegantly formulated in a temporal gauge that corresponds to the following choice \cite{Vignale2004}
\begin{equation}
 \label{t-gauge}
\chi(\xv,t) = \int_{0}^{t}U(\xv,t')dt'.
\end{equation}
The transformation with the gauge function (\ref{t-gauge}) eliminates the scalar potential so that the many-body Hamiltonian reduces to the form
\begin{equation} 
H[\Av] = \sum_{j = 1}^{N}\frac{(-i\nabla_{j} - {\bf A}(\xv_{j},t))^{2}}{2m} +
\frac{1}{2}\sum_{j\ne k} V(|\xv_{j}-\xv_{k}|).
\label{H-t-gauge}
\end{equation}

The key physical observables in TDCDFT are the density of particles $n(\xv,t)$ and the current density $\jv(\xv,t)$:
\begin{eqnarray}
\label{n-def}
n(\xv,t) &=& \langle\Psi(t)|\hat{n}(\xv)|\Psi(t)\rangle, \\
 \label{j-def}
\jv(\xv,t) &=& \langle\Psi(t)|\hat{\jv}^{p}(\xv)|\Psi(t)\rangle - \frac{n(\xv,t)}{m}\Av(\xv,t),
\end{eqnarray}
where $\Psi(t)$ is the solution to the Schr\"odinger equation (\ref{SE}), and $\hat{n}(\xv)$ and $\hat{\jv}^{p}(\xv)$ are the operators of the density and of the paramagnetic current, respectively,
\begin{eqnarray}
 \label{n-oper}
\hat{n}(\xv) &=& \sum_{j=1}^{N}\delta(\xv - \xv_j), \\
\label{jpar-oper}
\hat{\jv}^{p}(\xv) &=& \frac{-i}{2m}\sum_{j=1}^{N}\{\nabla_{j},\delta(\xv - \xv_j)\}.
\end{eqnarray}
The gauge invariance implies that the number of particles is locally conserved, i.~e. the density and the current are connected by the continuity equation
\begin{equation}
 \label{continuity}
\partial_t n(\xv,t) + \nabla\jv(\xv,t)=0.
\end{equation}
This, in particular, means that $n(\xv,t)$ is uniquely determined by the current $\jv(\xv,t)$ and the initial density distribution $n(\xv,0)=n_0(\xv)$.

The usual statement of the problem in quantum mechanics corresponds to solving the time-dependent Schr\"odinger equation (\ref{SE}), (\ref{H-t-gauge}), for a given initial condition (\ref{InitialPsi}) and the external driving potential $\Av(\xv,t)$. This determines the many-body wave function $\Psi[\Psi_0,\Av](t)$ and thus any observable, e.~g. the current $\jv[\Psi_0,\Av](\xv,t)$, as unique functionals of the initial state $\Psi_0$ and the vector potential $\Av$. The existence of TDCDFT assumes that the inverse problem also possess a unique solution. Namely, TDCDFT is valid if, given the initial state $\Psi_0$ and the current $\jv(\xv,t)$, one can uniquely reconstruct the time-dependent wave function $\Psi[\Psi_0,\jv](t)$ and the potential $\Av[\Psi_0,\jv](\xv,t)$ which supports the prescribed current. 

The key observation is that, by ``reinterpreting'' the definition (\ref{j-def}) of the current, the above inverse problem can be mathematically posed in a form of the following {\it nonlinear} system of equations
\begin{eqnarray}
 \label{NLSE-TDCDFT}
i\partial_t\Psi(t) &=& H[\Av]\Psi(t), \\
\label{A(j)}
\Av(\xv,t) &=& \frac{m}{n(\xv,t)}\left[\langle\Psi(t)|\hat{\jv}^{p}(\xv)|\Psi(t)\rangle - \jv(\xv,t)\right],
\end{eqnarray}
which has to be solved with the initial condition (\ref{InitialPsi}). The Hamiltonian in (\ref{NLSE-TDCDFT}) still has a form of Eq.~(\ref{H-t-gauge}). However, now the vector potential $\Av$ is not fixed externally, but calculated selfconsistently from Eq.~(\ref{A(j)}). The density $n(\xv,t)$ entering (\ref{A(j)}) is determined either by intergating the continuity equation (\ref{continuity}) or calculated directly as the expectation value of Eq.~(\ref{n-def}). Both definitions are identical if the wave function $\Psi(t)$ satisfies Eq.~(\ref{NLSE-TDCDFT}). By inserting the selfconsistent vector potential (\ref{A(j)}) into (\ref{NLSE-TDCDFT}) we obtain NLSE with a special cubic nonlinearity. The solution of the Cauchy problem for this NLSE, provided it is well posed, returns the time-dependent many-body wave function and the vector potential as universal functionals of the initial state $\Psi_0$ and the given current $\jv(\xv,t)$. Therefore the problem of existence of TDCDFT reduces to proving the existence and uniqueness of solutions to NLSE defined by Eqs.~(\ref{NLSE-TDCDFT}) and (\ref{A(j)}).

To illustrate how the NLSE approach works I consider the simplest case of one quantum particle. For $N=1$ the nonlinear Cauchy problem (\ref{NLSE-TDCDFT}), (\ref{A(j)}), and (\ref{InitialPsi}) simplifies as follows
\begin{eqnarray}
 \label{SE-1part}
i\partial_t\Psi(\xv,t) &=& \frac{1}{2m}(-i\nabla-\Av(\xv,t))^2\Psi(\xv,t),\\
\label{A(j)-1part}
\Av(\xv,t) &=& \frac{1}{|\Psi|^2}\left[\frac{-i}{2}(\Psi^*\nabla\Psi - \Psi\nabla\Psi^*)-m\jv(\xv,t)\right],\\
\label{Init-1part}
\Psi(\xv,0) &=& \Psi_0(\xv) \equiv \sqrt{n_0(\xv)}e^{i\varphi_0(\xv)}.
\end{eqnarray}
On the first sight this nonlinear problem may look complicated, but it turns out to be trivially integrable. The corresponding exact analytic solution can be represented in the form
\begin{eqnarray}
 \label{Psi-1part}
\Psi[\Psi_0,\jv](\xv,t) &=& \sqrt{n(\xv,t)}e^{i\varphi(\xv,t)},\\
\label{A-1part}
\Av[\Psi_0,\jv](\xv,t) &=& \nabla\varphi(\xv,t) - m \frac{\jv(\xv,t)}{n(\xv,t)},
\end{eqnarray}
where the functions $n(\xv,t)$ and $\varphi(\xv,t)$ are defined as follows
\begin{eqnarray}
 \label{n-1part}
n(\xv,t) &=& n_0(\xv) - \int_0^t dt' \nabla\jv(\xv,t'),\\
\label{phase-1part}
\varphi(\xv,t) &=& \varphi_0(\xv) + \int_0^t dt'\left[\frac{\nabla^2\sqrt{n(\xv,t')}}{2m\sqrt{n(\xv,t')}} -
\frac{m\jv^2(\xv,t')}{2n^2(\xv,t')}\right].
\end{eqnarray}
Equations (\ref{Psi-1part})--(\ref{phase-1part}) provide us with an explicit example of the universal TDCDFT functionals $\Psi[\Psi_0,\jv](t)$ and $\Av[\Psi_0,\jv](\xv,t)$ recovered from the NLSE (\ref{NLSE-TDCDFT}), (\ref{A(j)}) for a particular case of $N=1$. 

Apparently the general NLSE for any $N>1$ is not solvable analytically. However the mathematical structure of equations remains practically the same, which leaves a strong hope that early or later the uniqueness and existence theorems for the general nonlinear many-body problem, and thus the existence of TDCDFT, will be proved. In fact, recently I have succeeded to prove the corresponding theorems for a lattice TDCDFT \cite{Tokatly2010arxiv} which is formulated in terms of a discrete (actually finite-difference) version of the above NLSE.

\subsection{Equivalence of the nonlinear universal problems in TDCDFT and TDDefFT}
The NLSE formulation has been first formulated in the context of TDDefFT where it appears as a natural ``universal'' step in solving the many-body in the comoving reference frame \cite{TokatlyPRB2007,TokatlyPCCP2009}. Below I will show that there is a one-to-one correspondence between the universal problem derived in Refs.~\cite{TokatlyPRB2007,TokatlyPCCP2009} for TDDefFT and the NLSE (\ref{NLSE-TDCDFT}), (\ref{A(j)}) for TDCDFT. To prove this correspondence we proceed as follows. First, for a given current $\jv(\xv,t)$, we integrate the continuity equation (\ref{continuity}) to calculate the density $n(\xv,t)$, and then define the velocity field:
\begin{equation}
 \label{velocity}
\vv(\xv,t) = \frac{\jv(\xv,t)}{n(\xv,t)}.
\end{equation}
Using the velocity field $\vv(\xv,t)$ one finds a set of Lagrangian trajectories $\xv(\xiv,t)$ by solving the following initial value problem (see, e.~g. Refs.~\cite{TokatlyPRB2007,TokatlyPCCP2009})
\begin{equation}
 \label{trajectory}
\dot\xv(\xiv,t) = \vv(\xv(\xiv,t),t)\, \qquad \xv(\xiv,0) = \xiv,
\end{equation}
where $\dot\xv(\xiv,t)=\partial_t\xv(\xiv,t)$. Physically the function $\xv(\xiv,t)$ is a trajectory of an infinitesimal fluid element which start its motion at $t=0$ from the point $\xiv$. The formal significance of the trajectory function $\xv(\xiv,t)$ is that the equation $\xv=\xv(\xiv,t)$ defines a transformation of coordinates $\xv\to\xiv$ which corresponds the transformation from the laboratory frame ($\xv$-space) to the comoving Lagrangian frame ($\xiv$-space). 

The last step in proving the equivalence of NLSE in TDCDFT and TDDefFT is to rewrite Eqs.~(\ref{NLSE-TDCDFT}) and (\ref{A(j)}) in the comoving frame. Formally we perform the transformation of coordinates $\xv\to\xiv$ : $\xv=\xv(\xiv,t)$ and define the transformed many-body wave function $\widetilde{\Psi}(\xiv_1,\dots,\xiv_N,t)$ in the comoving frame as follows \cite{TokatlyPRB2007,TokatlyPCCP2009}
\begin{equation}
\widetilde{\Psi}({\bm\xi}_{1},\dots,{\bm\xi}_{N},t)
= \prod_{j = 1}^{N}g^{\frac{1}{4}}({\bm\xi}_{j},t)
e^{-iS_{\text{cl}}({\bm\xi}_{j},t)}\Psi({\bf x}({\bm\xi}_{1},t),\dots,{\bf x}({\bm\xi}_{N},t),t),
\label{Psi-Lagr}
\end{equation}
where $g(\xiv,t)={\rm det}[g_{\mu\nu}(\xiv,t)]$ is the determinant of the metric tensor induced by the transformation of coordinates ($\sqrt{g}$ is the Jacobian of the transformation)
\begin{equation}
 \label{metric}
g_{\mu\nu}(\xiv,t) = \frac{\partial x^{\alpha}}{\partial\xi^{\mu}}
\frac{\partial x^{\alpha}}{\partial\xi^{\mu}}; \quad
g^{\mu\nu}(\xiv,t) = [g_{\mu\nu}]^{-1} = \frac{\partial\xi^{\mu}}{\partial x^{\alpha}}
\frac{\partial\xi^{\nu}}{\partial x^{\alpha}},
\end{equation}
and $S_{\text{cl}}({\bm\xi},t)$ is the classical action of a particle moving along the trajectory ${\bf x}(\bm\xi,t)$
\begin{equation}
\label{action}
S_{\text{cl}}({\bm\xi},t)= \int_0^t\left[ 
\frac{m}{2}(\dot{\bf x}(\xiv,t))^{2} 
+ \dot{\bf x}(\xiv,t){\bf A}({\bf x}(\xiv,t),t)\right].
\end{equation}
Note that the factor $\prod_{j = 1}^{N}g^{\frac{1}{4}}({\bm\xi}_{j},t)$ in (\ref{Psi-Lagr}) ensures the standard normalization of the wave function $\langle\widetilde{\Psi}|\widetilde{\Psi}\rangle=1$ after a non-volume-preserving transformation of coordinates. After straightforward algebra we find that the transformed Schr\"odinger equation (\ref{NLSE-TDCDFT}) and the ``self-consistency'' equation (\ref{A(j)}) can be reduced to the following form
\begin{eqnarray}
 \label{NLSE-DFDefFT}
i\partial_t\widetilde{\Psi}({\bm\xi}_{1},\dots,{\bm\xi}_{N},t) &=&
\widetilde{H}[g_{ij},\bm{\mathcal A}]\widetilde{\Psi}({\bm\xi}_{1},\dots,{\bm\xi}_{N},t),\\
\label{A(g)}
\Acalv(\xiv,t) &=& \frac{m}{n_0(\xiv)}\langle\widetilde{\Psi}(t)|\hat{\jv}^{p}(\xiv)|\widetilde{\Psi}(t)\rangle.
\end{eqnarray}
In Eq.~(\ref{NLSE-DFDefFT}) $\widetilde{H}[g_{ij},\bm{\mathcal A}]$ is the Hamiltonian (\ref{H-t-gauge}) transformed to the comoving frame:
\begin{equation}
 \label{H-Lagr}
\widetilde{H}[g_{ij},\bm{\mathcal A}] = \sum_{j = 1}^{N}g^{-\frac{1}{4}}_{j}
\hat{K}_{j,\mu}\frac{\sqrt{g_{j}}g^{\mu\nu}_{j}}{2m}
\hat{K}_{j,\nu}g^{-\frac{1}{4}}_{j}
+\frac{1}{2}\sum_{k\ne j}V(l_{\bm\xi_{k}\bm\xi_{j}})
\end{equation}
where $\hat{K}_{j,\mu}=-i\partial_{\xi^{\mu}_{j}}
- {\cal A}_{\mu}(\bm\xi_{j},t)$, $g^{\mu\nu}_{j}=g^{\mu\nu}(\bm\xi_{j},t)$, and $l_{\bm\xi_{k}\bm\xi_{j}}$ is the distance between $j$th and $k$th particles in the moving frame (the length of geodesic connecting points $\bm\xi_{j}$ and $\bm\xi_{k}$ in the space with metric $g_{\mu\nu}$). The selfconsistent vector potential $\Acalv(\xiv,t)$ in (\ref{NLSE-DFDefFT}), (\ref{A(g)}) is related to $\Av(\xv,t)$ entering (\ref{NLSE-TDCDFT}) and (\ref{A(j)}) as follows
 \begin{equation}
 \label{A-TDDefFT}
\Acal_{\mu}(\xiv,t) = \frac{\partial x^{\nu}}{\partial\xi^{\mu}}A_{\nu}(\xv(\xiv,t),t) 
+ \frac{\partial x^{\nu}}{\partial\xi^{\mu}}\dot{x}^{\nu} - \partial_{\xi^{\mu}}S_{\text{cl}}({\bm\xi},t).
\end{equation}
The first term in the right hand side in this equation is the transformed left hand side in (\ref{A(j)}). The second term originates from the term $m\jv/n$ in (\ref{A(j)}), while the last term in (\ref{A-TDDefFT}) comes from the expectation value of the paramagnetic current in (\ref{A(j)}) and is related to the extra phase factor in the transformed wave function (\ref{Psi-Lagr}).

The nonlinear system of equations (\ref{NLSE-DFDefFT}), (\ref{A(g)}) is exactly the universal many-body problem formulated within TDDefFT (see Eqs.~(58), (59) in \cite{TokatlyPRB2007} or Eqs.~(3.38), (3.39) in \cite{TokatlyPCCP2009}). In fact, the NLSE in TDCDFT and TDDefFT correspond to the same mathematical problem formulated in the laboratory or the comoving frame respectively. Hence to prove the existence of TDCDFT we are free to choose any of the two formulations. The advantage of the comoving frame formulation is a much simpler form of the selfconsistency equation. Indeed Eq.~(\ref{A(g)}) does not contain an inhomogeneous term and involves only the initial density distribution $n_0$. The price for that -- the appearance of the time-dependent metric tensor in the Schr\"odinger equation -- is probably not too high from the conceptual point of view. On the other hand, the formulation derived in this paper, Eqs.~ (\ref{NLSE-TDCDFT}) and (\ref{A(j)}), is much easier to digest as it is based on the usual quantum dynamics in the laboratory frame. 

\subsection{Connection of the nonlinear inverse problem in TDCDFT to the Vignale proof of the mapping theorem}

In Ref.~\cite{Vignale2004} Vignale generalized the van~Leeuwen construction \cite{vanLeeuwen1999} to prove the mapping theorem for TDCDFT. The main idea of the proof is to relate the vector potential $\Av(\xv,t)$ to the current and the many-body wave function using the force balance equation. Then this relation has been interpreted as an equation of motion for $\Av(\xv,t)$ and solved recursively assuming $t$-analyticity of the vector potential. The present subsection is aimed at demonstrating that the statement the problem used in \cite{Vignale2004} directly follows from the nonlinear system of Eqs.~(\ref{NLSE-TDCDFT}) and (\ref{A(j)}).

Let us first multiply both sides of (\ref{A(j)}) with the density $n(\xv,t)$ and differentiate the result with respect to time:
\begin{equation}
 \label{dA(j)/dt}
n\partial_t\Av + \Av\partial_t n = m\partial_t\langle\Psi(t)|\hat{\jv}^{p}(\xv)|\Psi(t)\rangle 
- m\partial_t\jv .
\end{equation}
The time derivative of the paramagnetic current $\jv^{p}(\xv,t)\equiv\langle\Psi(t)|\hat{\jv}^{p}(\xv)|\Psi(t)\rangle$ in the right hand side of (\ref{dA(j)/dt}) can be straightforwardly calculated using the Schr\"odinger equation (\ref{NLSE-TDCDFT}). The result of the differentiation takes the following form
\begin{equation}
 \label{djp/dt}
 m\partial_t j_{\mu}^p = A_{\mu}\partial_t n + \left[\left(\jv^{p} - \frac{n}{m}\Av\right)\times\Bv\right]_{\mu}
- \partial_{\nu}\Pi_{\mu\nu},
\end{equation}
where $\Bv=\nabla\times\Av$ is the magnetic field associated with the selfconsistent vector potential, and the last term is the stress force with $\Pi_{\mu\nu}(\xv,t)=\langle\Psi(t)|\hat{\Pi}_{\mu\nu}(\xv)|\Psi(t)\rangle$ being the stress tensor. An explicit form of the stress force and the stress tensor can be found, for example, in \cite{Vignale2004,TokatlyPRB2005a,TokatlyPCCP2009}. Inserting the result of (\ref{djp/dt}) into (\ref{dA(j)/dt}) we arrive at the following equation of motion for the vector potential
\begin{equation}
 \label{force-balance1}
n\partial_t A_{\mu} = \left[\left(\jv^{p} - \frac{n}{m}\Av\right)\times\Bv\right]_{\mu} - \partial_{\nu}\Pi_{\mu\nu}
- m\partial_t j_{\mu},
\end{equation}
which is exactly the force balance equation. Using the definition of (\ref{j-def}), which is an absolutely legitimate operation at the solution point, we find that (\ref{force-balance1}) becomes identical to Eq.~(10) in \cite{Vignale2004}. Hence the force balance equation, which has been used in \cite{Vignale2004} to construct a recursion for $\Av[\jv](t)$, is nothing but the time derivative of the selfconsistency equation (\ref{A(j)}). In other words, the force balance equation, considered as the equation of motion for the vector potential $\Av(\xv,t)$, can be integrated explicitly and its solution is given by (\ref{A(j)}).

Apparently the equation for $\Av$ (\ref{A(j)}) is advantageous as it leads to a much simpler and cleaner form of the nonlinearity in the inverse many-body problem. Another important point is that Eq.~(\ref{A(j)}) is local in time. Just from the structure of the nonlinear problem (\ref{NLSE-TDCDFT}), (\ref{A(j)}) we immediately see that, if the solution exists, the vector potential $\Av[\jv](t)$ is a strictly retarded functional of the current $\jv(\xv,t)$. In particular, it can not contain even local terms which depend on the time derivative of the current. The time locality of (\ref{A(j)}) also makes it absolutely obvious that the power series expansion of $\Av(t)$ is unique, which is the essence of the Vignale theorem \cite{Vignale2004}. Of course, to construct such an expansion we have to assume, as usual, the $t$-analyticity of the current. I am not presenting here the explicit construction of the recursion because it is practically trivial, and basically reproduces the recursion for the selfconsistent potential $\Acalv(t)$ in the universal problem (\ref{NLSE-DFDefFT}), (\ref{A(g)}) for TDDefFT \cite{TokatlyPRB2007,TokatlyPCCP2009}. Hence the present derivation of the nonlinear inverse problem (\ref{NLSE-TDCDFT}), (\ref{A(j)}) can be also viewed as a considerable simplification of the proof for the Vignale theorem. Clearly the convergence of the series remains unproved as in all power series based approaches to the problem of TDDFT and TDCDFT \cite{vanLeeuwen1999,Vignale2004,TokatlyPRB2007}. The hope is that a cleanly stated nonlinear initial value problem, (\ref{NLSE-TDCDFT}), (\ref{A(j)}), can be treated using more advanced mathematical tools as it has been done recently for a lattice version of the theory \cite{Tokatly2010arxiv}.

\section{Time-dependent density functional theory}

In Sec.~II we have found that the problem of existence of TDCDFT is equivalent to that of existence and uniqueness of a solution to a certain time-dependent NLSE. We have proved a one-to-one correspondence of this NLSE to the universal nonlinear problem derived earlier for TDDefFT, and demonstrated its close relation to the Vignale's power-series proof of the TDCDFT mapping theorem. Recently a conceptually similar NLSE setup has been also proposed for TDDFT \cite{Maitra2010}. In this section I will derive an alternative formulation of the TDDFT problem, and establish a connection between the results of Sec.~II (i.~e., the NLSE formulation of TDDefFT and TDCDFT) to the NLSE proposed in Ref.~\cite{Maitra2010}.

The main statement of TDDFT is that the density $n(\xv,t)$ completely determines the many-body dynamics, provided the latter is driven by a scalar potential. Hence our starting point is the many-body Schr\"odinger equation (\ref{SE}) with a Hamiltonian that contains only the external scalar potential as a driving force:
\begin{equation} 
H = \sum_{j = 1}^{N}\left[-\frac{\nabla_{j}^2}{2m} + U(\xv_{j},t)\right] +
\frac{1}{2}\sum_{j\ne k} V(|\xv_{j}-\xv_{k}|).
\label{H-U}
\end{equation}
At the formal level this standard form of the Hamiltonian assumes that a given, purely longitudinal external field is described using a Coulomb gauge ($\nabla\Av =0$). 

Following the ideas of the previous section we perform a gauge transformation (\ref{Psi-gauge}) with the gauge function $\chi(\xv,t)$ defined by Eq.~(\ref{t-gauge}). In other words, we switch from the original Coulomb gauge to a temporal gauge, where the scalar potential is absent, but instead the dynamics is driven by a purely longitudinal vector potential $\Av(\xv,t)=\nabla\chi(\xv,t)$. The Hamiltonian in the temporal gauge takes the form
\begin{equation} 
H[\chi] = \sum_{j = 1}^{N}\frac{[-i\nabla_{j} - \nabla_{j}\chi(\xv_{j},t)]^{2}}{2m} +
\frac{1}{2}\sum_{j\ne k} V(|\xv_{j}-\xv_{k}|).
\label{H-chi}
\end{equation}
Obviously, the driving force is still determined by a single scalar function $\chi(\xv,t)$ which equals to the potential $U(\xv,t)$ integrated over the time, Eq.~(\ref{t-gauge}). 

Solution of the Schr\"odinger with the Hamiltonian $H[\chi]$ yields the many-body wave function $\Psi[\Psi_0,\chi](t)$ which can be used to calculate any observable, for example the density or the current, as a functional of the initial state $\Psi_0$ and the driving ``potential'' $\chi(\xv,t)$. It is worth noting that the density $n(\xv,t)$ is always given by the gauge independent Eq.~(\ref{n-def}), while the formal expression for the current $\jv(\xv,t)$ in the temporal gauge acquires a diamagnetic contribution:
\begin{equation}
 \label{j(chi)}
\jv(\xv,t) = \langle\Psi(t)|\hat{\jv}^{p}(\xv)|\Psi(t)\rangle - \frac{n(\xv,t)}{m}\nabla\chi(\xv,t)
\end{equation}

For the discussion below the following simple observation is of primary importance. By taking the divergence of Eq.~(\ref{j(chi)}) and using the continuity equation (\ref{continuity}) we find that the time derivative of the density, $\partial_t n(\xv,t)$, can be calculated directly as the following expectation value
\begin{equation}
 \label{dn(chi)}
\partial_t n(\xv,t) = -\langle\Psi(t)|\nabla\hat{\jv}^{p}(\xv)|\Psi(t)\rangle +
 \frac{1}{m}\nabla[n(\xv,t)\nabla\chi(\xv,t)].
\end{equation}
The relevance of this equation for TDDFT should be obvious from the familiar looking Sturm-Liouville operator \cite{vanLeeuwen1999} in the right hand side (the second term).

The validity of TDDFT means that, given the initial state $\Psi_0$ and the density $n(\xv,t)$, we can uniquely recover the time-dependent many-body wave function $\Psi[\Psi_0,n](t)$ and the corresponding potential which gives rise to the prescribed density. In the temporal gauge this inverse problem can be formulated in terms of the following nonlinear system of equations
\begin{eqnarray}
 \label{NLSE-TDDFT}
i\partial_t\Psi(t) &=& H[\chi]\Psi(t), \\
\label{chi(n)}
\nabla[n(\xv,t)\nabla\chi(\xv,t)] &=& m\langle\Psi(t)|\nabla\hat{\jv}^{p}(\xv)|\Psi(t)\rangle,
+m\partial_t n(\xv,t).
\end{eqnarray}
where Eq.~(\ref{chi(n)}) corresponds to Eq.~(\ref{dn(chi)}) reinterpreted as an equation for $\chi(\xv,t)$. The system (\ref{NLSE-TDDFT}), (\ref{chi(n)}) has to be supplemented with the initial condition $\Psi(0)=\Psi_0$, and solved for a given $n(\xv,t)$.

The Hamiltonian in the Schr\"odinger equation (\ref{NLSE-TDDFT}) is the same as in the usual linear quantum problem, i.~e. it is defined by Eq.~(\ref{H-chi}). However, the ``potential'' function $\chi(\xv,t)$ in $H[\chi]$ is determined selfconsistently  from Eq.~(\ref{chi(n)}) for a given density. Mathematically the system of (\ref{NLSE-TDDFT}), (\ref{chi(n)}) is equivalent to NLSE with a spatially nonlocal cubic nonlinearity. Provided the Cauchy problem for this NLSE is well posed, its solution returns the functionals $\Psi[\Psi_0,n](t)$ and $\chi[\Psi_0,n](\xv,t)$. A functional, which corresponds to the scalar potential in the original gauge, is calculated as $U[\Psi_0,n](\xv,t) = \partial_t\chi[\Psi_0,n](\xv,t)$ according to the definition (\ref{t-gauge}) of the gauge function. Hence the proof of the density-potential mapping and the $\Vcal$-representability reduces to proving the uniqueness and the existence of solutions for the nonlinear problem (\ref{NLSE-TDDFT}), (\ref{chi(n)}). In this setting the existence corresponds to the $\Vcal$-representability, while the uniqueness of a solution is equivalent to the Runge-Gross mapping theorem.

A connection of the above NLSE for TDDFT to the corresponding problem in TDCDFT (see Sec.~II) is practically obvious. Indeed Eq.~(\ref{chi(n)}) is nothing but a divergence of Eq.~(\ref{A(j)}) when the latter is restricted to longitudinal vector potentials of the form $\Av=\nabla\chi$. Hence TDDFT can be considered as a particular case of TDCDFT for systems driven by a purely longitudinal vector potential. Since in this case the vector potential is uniquely determined by a single scalar function $\chi(\xv,t)$, a knowledge of a single scalar collective variable, the density $n(\xv,t)$, is sufficient to uniquely recover both the full many-body wave function and the driving potential. 

Let us now demonstrate the equivalence of Eqs.~(\ref{NLSE-TDDFT}), (\ref{chi(n)}) to the NLSE formulation proposed in \cite{Maitra2010}. The argumentation below is very similar to that used in Sec.~II to prove the equivalence of our NLSE formulation (\ref{NLSE-TDCDFT}), (\ref{A(j)}), and the Vignale's approach to the TDCDFT mapping problem. As a first step we differentiate Eq.~(\ref{chi(n)}) with respect to $t$:
\begin{equation}
 \label{dchi/dt}
\nabla[\dot{n}\nabla\chi] + \nabla[n\nabla\dot{\chi}] = m\partial_t\langle\Psi(t)|\nabla\hat{\jv}^{p}(\xv)|\Psi(t)\rangle
+m\ddot{n}.
\end{equation}
Next, we use the Schr\"odinger equation (\ref{NLSE-TDDFT}) to compute the time derivative of the quantity $\langle\Psi(t)|\nabla\hat{\jv}^{p}(\xv)|\Psi(t)\rangle$ entering the right hand side in (\ref{dchi/dt}). This is equivalent to simply setting $\Av=\nabla\chi$ in Eq.~(\ref{djp/dt}) and taking its divergence, which yields the following result
\begin{equation}
 \label{divdjp/td}
m\partial_t\langle\Psi(t)|\nabla\hat{\jv}^{p}(\xv)|\Psi(t)\rangle = \nabla[\dot{n}\nabla\chi]
-\partial_{\mu}\partial_{\nu}\Pi_{\mu\nu}.
\end{equation}
Inserting (\ref{divdjp/td}) into (\ref{dchi/dt}) we obtain the following form of the selfconsistency equation,
\begin{equation}
 \label{force-balance2}
\nabla[n(\xv,t)\nabla U(\xv,t)]= m\ddot{n}(\xv,t)-\partial_{\mu}\partial_{\nu}\Pi_{\mu\nu}(\xv,t),
\end{equation}
where $U(\xv,t)\equiv\partial_t\chi(\xv,t)$ [see Eq.~(\ref{t-gauge})] is the usual scalar potential. Apparently Eq.~(\ref{force-balance2}) is equivalent to the equation (5) in Ref.~\cite{Maitra2010}. More precisely, after the transformation to the Coulomb gauge, equations (\ref{force-balance2}) and (\ref{NLSE-TDDFT}) identically reproduce equations (5) and (6) of Ref.~\cite{Maitra2010}. This proves the equivalence of the present formulation of TDDFT problem and NLSE proposed in \cite{Maitra2010}. 

Now we can answer the question raised by the authors of Ref.~\cite{Maitra2010} -- what is the relation between the particular NLSE derived for TDDefFT in \cite{TokatlyPRB2007,TokatlyPCCP2009}, and the one proposed for TDDFT in \cite{Maitra2010}. Starting from the TDDefFT universal problem, (\ref{NLSE-DFDefFT}) and (\ref{A(g)}), we successively perform the following transformations. (i) First, we transform the TDDefFT-NLSE of Eqs.~(\ref{NLSE-DFDefFT}), (\ref{A(g)}) from the comoving to the laboratory frame. The result is the TDCDFT-NLSE, (\ref{NLSE-TDCDFT}), (\ref{A(j)}). (ii) Then, in the TDCDFT framework we consider only longitudinal vector potentials by setting $\Av=\nabla\chi$ in (\ref{NLSE-TDCDFT}), (\ref{A(j)}), and taking the divergence of Eq.~(\ref{A(j)}). This brings us to the TDDFT-NLSE in the form (\ref{NLSE-TDDFT}), (\ref{chi(n)}). (iii) Next, we differentiate (\ref{chi(n)}) with respect to time and, as a result, arrive at the TDDFT-NLSE of (\ref{NLSE-TDDFT}), (\ref{force-balance2}). (iv) Finally, we switch from the temporal to the Coulomb gauge, and transform Eqs.~(\ref{NLSE-TDDFT}) and (\ref{force-balance2}) to the TDDFT-NLSE proposed in Ref.~\cite{Maitra2010}.

\section{Conclusion}

The main result of the present work is a unified approach to the density-potential mapping and the $\Vcal$-representability problems in TDDFT, TDCDFT, and TDDefFT. The existence of each theory in the above family is mapped to the uniqueness and existence of solutions to a certain nonlinear many-body problem. We have uncovered close relations between the nonlinear problems for all three theories, and established links to all previously known approaches to TDDFT and TDCDFT. The analysis performed in this work shows a special role of TDCDFT in the whole problematics of the TDDFT-like formalisms. Indeed, starting the nonlinear universal problem, (\ref{NLSE-TDCDFT}), (\ref{A(j)}), for TDCDFT one can easily generate the corresponding nonlinear problems for the other members of the family. The NLSE formulation of TDCDFT demonstrates the main features of the new approach in the most transparent form. It is free of the additional Sturm-Liouville-type nonlocality of TDDFT, and does not involve complications related to the time-dependent metrics as TDDefFT. This clearly indicate that first it makes a sense to concentrate on the existence and uniqueness theorems for the TDCDFT nonlinear problem of Eqs.~(\ref{NLSE-TDCDFT}), (\ref{A(j)}). Provided this problem is solved the corresponding results for the other members of the family should follow almost automatically.

\bigskip
This work was supported by the Spanish MICINN (FIS2010-21282-C02-01), ``Grupos Consolidados UPV/EHU del Gobierno
Vasco'' (IT-319-07), and the European Union through e-I3 ETSF project (Contract No. 211956).

%\bibliography{journals,bose,kinetics,tddft,DFT,hydrodynamics,geometry,mypapers,books,SpinOrbit,Graphene,NLSE-TDDFT_note}

\begin{thebibliography}{13}
\expandafter\ifx\csname natexlab\endcsname\relax\def\natexlab#1{#1}\fi
\expandafter\ifx\csname bibnamefont\endcsname\relax
  \def\bibnamefont#1{#1}\fi
\expandafter\ifx\csname bibfnamefont\endcsname\relax
  \def\bibfnamefont#1{#1}\fi
\expandafter\ifx\csname citenamefont\endcsname\relax
  \def\citenamefont#1{#1}\fi
\expandafter\ifx\csname url\endcsname\relax
  \def\url#1{\texttt{#1}}\fi
\expandafter\ifx\csname urlprefix\endcsname\relax\def\urlprefix{URL }\fi
\providecommand{\bibinfo}[2]{#2}
\providecommand{\eprint}[2][]{\url{#2}}

\bibitem[{\citenamefont{Marques et~al.}(2006)\citenamefont{Marques, Noguiera,
  Rubio, Burke, Ullrich, and Gross}}]{TDDFT2006}
\bibinfo{editor}{\bibfnamefont{M.~A.~L.} \bibnamefont{Marques}},
  \bibinfo{editor}{\bibfnamefont{F.}~\bibnamefont{Noguiera}},
  \bibinfo{editor}{\bibfnamefont{A.}~\bibnamefont{Rubio}},
  \bibinfo{editor}{\bibfnamefont{K.}~\bibnamefont{Burke}},
  \bibinfo{editor}{\bibfnamefont{C.~A.} \bibnamefont{Ullrich}},
  \bibnamefont{and} \bibinfo{editor}{\bibfnamefont{E.~K.~U.}
  \bibnamefont{Gross}}, eds., \emph{\bibinfo{title}{Time-Dependent Density
  Functional Theory}}, vol. \bibinfo{volume}{706} of
  \emph{\bibinfo{series}{Lecture Notes in Physics}}
  (\bibinfo{publisher}{Springer}, \bibinfo{address}{Berlin},
  \bibinfo{year}{2006}).

\bibitem[{\citenamefont{Marques and Rubio}(2009)}]{TDDFT-PCCP}
\bibinfo{author}{\bibfnamefont{M.~A.~L.} \bibnamefont{Marques}}
  \bibnamefont{and} \bibinfo{author}{\bibfnamefont{A.}~\bibnamefont{Rubio}},
  \bibinfo{journal}{Phys. Chem. Chem. Phys.} \textbf{\bibinfo{volume}{11}},
  \bibinfo{pages}{4421} (\bibinfo{year}{2009}).

\bibitem[{\citenamefont{Casida et~al.}(2009)\citenamefont{Casida, Chermette,
  and Jacquemin}}]{TDDFT-TEOCHEM}
\bibinfo{author}{\bibfnamefont{M.~E.} \bibnamefont{Casida}},
  \bibinfo{author}{\bibfnamefont{H.}~\bibnamefont{Chermette}},
  \bibnamefont{and}
  \bibinfo{author}{\bibfnamefont{D.}~\bibnamefont{Jacquemin}},
  \bibinfo{journal}{Journal of Molecular Structure: THEOCHEM}
  \textbf{\bibinfo{volume}{914}}, \bibinfo{pages}{1 } (\bibinfo{year}{2009}).

\bibitem[{\citenamefont{Runge and Gross}(1984)}]{RunGro1984}
\bibinfo{author}{\bibfnamefont{E.}~\bibnamefont{Runge}} \bibnamefont{and}
  \bibinfo{author}{\bibfnamefont{E.~K.~U.} \bibnamefont{Gross}},
  \bibinfo{journal}{Phys. Rev. Lett.} \textbf{\bibinfo{volume}{52}},
  \bibinfo{pages}{997} (\bibinfo{year}{1984}).

\bibitem[{\citenamefont{van Leeuwen}(1999)}]{vanLeeuwen1999}
\bibinfo{author}{\bibfnamefont{R.}~\bibnamefont{van Leeuwen}},
  \bibinfo{journal}{Phys. Rev. Lett.} \textbf{\bibinfo{volume}{82}},
  \bibinfo{pages}{3863} (\bibinfo{year}{1999}).

\bibitem[{\citenamefont{Ghosh and Dhara}(1988)}]{GhoDha1988}
\bibinfo{author}{\bibfnamefont{S.~K.} \bibnamefont{Ghosh}} \bibnamefont{and}
  \bibinfo{author}{\bibfnamefont{A.~K.} \bibnamefont{Dhara}},
  \bibinfo{journal}{Phys. Rev. A} \textbf{\bibinfo{volume}{38}},
  \bibinfo{pages}{1149} (\bibinfo{year}{1988}).

\bibitem[{\citenamefont{Vignale}(2004)}]{Vignale2004}
\bibinfo{author}{\bibfnamefont{G.}~\bibnamefont{Vignale}},
  \bibinfo{journal}{Phys. Rev. B} \textbf{\bibinfo{volume}{70}},
  \bibinfo{pages}{201102(R)} (\bibinfo{year}{2004}).

\bibitem[{\citenamefont{Maitra et~al.}(2010)\citenamefont{Maitra, Todorov,
  Woodward, and Burke}}]{Maitra2010}
\bibinfo{author}{\bibfnamefont{N.~T.} \bibnamefont{Maitra}},
  \bibinfo{author}{\bibfnamefont{T.~N.} \bibnamefont{Todorov}},
  \bibinfo{author}{\bibfnamefont{C.}~\bibnamefont{Woodward}}, \bibnamefont{and}
  \bibinfo{author}{\bibfnamefont{K.}~\bibnamefont{Burke}},
  \bibinfo{journal}{Phys. Rev. A} \textbf{\bibinfo{volume}{81}},
  \bibinfo{pages}{042525} (\bibinfo{year}{2010}).

\bibitem[{\citenamefont{Tokatly}(2007)}]{TokatlyPRB2007}
\bibinfo{author}{\bibfnamefont{I.~V.} \bibnamefont{Tokatly}},
  \bibinfo{journal}{Phys. Rev. B} \textbf{\bibinfo{volume}{75}},
  \bibinfo{pages}{125105} (\bibinfo{year}{2007}).

\bibitem[{\citenamefont{Tokatly}(2009)}]{TokatlyPCCP2009}
\bibinfo{author}{\bibfnamefont{I.~V.} \bibnamefont{Tokatly}},
  \bibinfo{journal}{Phys. Chem. Chem. Phys.} \textbf{\bibinfo{volume}{11}},
  \bibinfo{pages}{4621 } (\bibinfo{year}{2009}).

\bibitem[{\citenamefont{Tokatly}(2010)}]{Tokatly2010arxiv}
\bibinfo{author}{\bibfnamefont{I.~V.} \bibnamefont{Tokatly}},
  \bibinfo{journal}{arXiv:1011.2715}  (\bibinfo{year}{2010}).

\bibitem[{\citenamefont{Ruggenthaler and {van
  Leeuwen}}(2010)}]{RugLee2010arxiv}
\bibinfo{author}{\bibfnamefont{M.}~\bibnamefont{Ruggenthaler}}
  \bibnamefont{and} \bibinfo{author}{\bibfnamefont{R.}~\bibnamefont{{van
  Leeuwen}}}, \bibinfo{journal}{arXiv:1011.3375}  (\bibinfo{year}{2010}).

\bibitem[{\citenamefont{Tokatly}(2005)}]{TokatlyPRB2005a}
\bibinfo{author}{\bibfnamefont{I.~V.} \bibnamefont{Tokatly}},
  \bibinfo{journal}{Phys. Rev. B} \textbf{\bibinfo{volume}{71}},
  \bibinfo{pages}{165104} (\bibinfo{year}{2005}).

\end{thebibliography}

%%%%%%%%%%%%%%%%%%%%%%%%%%%%%%%%%%%%%%%%%%%%%
\end{document}